
\input phyzzx
\hoffset=0.2truein
\voffset=0.1truein
\hsize=6truein
\def\TITLEPAGE{\frontpagetrue}
\def\CALT#1{\hbox to\hsize{\tenpoint \baselineskip=12pt
        \hfil\vtop{
        \hbox{\strut CALT-68-#1}}}}

\def\CALTECH{
        \address{California Institute of Technology,
Pasadena, CA 91125}}

\def\AUTHOR#1{\vskip .2in \centerline{#1}}

\def\ABSTRACT#1{\vskip .2in \vfil \centerline{\twelvepoint
\bf Abstract}
        #1 \vfil}
\def\ENDTITLEPAGE{\vfil\eject\pageno=1}

\def\Dslash{/\!\!\!\!D}
\def\vslash{/\!\!\!v}

\TITLEPAGE
\CALT{1963}
\bigskip            
\titlestyle {Heavy Quark Theory\foot{Work supported in part by the U.S. Dept.
of Energy
under Grant No. DE-FG03-92-ER40701.}}
\AUTHOR{Mark B. Wise}
\CALTECH

\singlespace
\bigskip
\centerline{{\it Talk presented at the Tennessee Int. Symposium on Radiative
Corrections:}}
\centerline{{\it Status and Outlook, Gatlinburg, TN, June 1994.}}
\bigskip
\normalspace

\ABSTRACT{Recent progress in the theory of hadrons containing a single heavy
quark is reviewed.  Particular attention is paid to those aspects that bear on
the determination of the magnitudes of the Cabibbo--Kobayashi--Maskawa matrix
elements $V_{cb}$ and $V_{ub}$.}

\ENDTITLEPAGE

\eject

\noindent{\bf 1.  Introduction}

Over the past year there have been several important developments in the theory
of hadrons containing a single heavy quark.  At the same time there have been
significant improvements from experiment in our understanding of the properties
of hadrons containing a charm or bottom quark.

The minimal standard model has six quarks that couple to the charged $W$-bosons
through the term
 $$
{\cal L}_{int} = {g_2\over 2\sqrt{2}} (\bar{u}, \bar{c}, \bar{t}) \gamma_\mu (1
- \gamma_5) V \left(\matrix{d \cr s \cr b\cr} \right) W^\mu + h.c.\eqno (1)
 $$
in the Lagrange density.  Here $g_2$ is the weak SU(2) coupling, $W^\mu$ is the
charged $W$-boson field and $V$ is the Cabibbo--Kobayashi--Maskawa matrix.  $V$
arises from the diagonalization of the quark mass matrices.  It can be written
in terms of three Euler like angles and a complex phase $e^{i\delta}$.  In the
minimal standard model it is this phase that is responsible for the CP
violation observed in kaon decay and CP violation in $B$ decay.  Extensions of
the standard model with extended Higgs sectors usually have additional sources
of CP violation.  It is hoped to test the correctness of the minimal standard
model for CP violation in future $B$ decay experiments and elsewhere.

In the minimal standard model the elements of the Cabibbo--Kobayashi--Maskawa
matrix are fundamental parameters that must be determined from experiment.
In this talk I will concentrate on those issues in heavy quark theory that are
related to a determination of $|V_{ub}|$ and $|V_{cb}|$ from $B$ decays.  Other
interesting areas where progress has occurred will, for the most part, be
omitted.   Even within the area of those elements of heavy quark physics
related to determining the weak mixing angles I will not be able to give a
complete review.  For example, I will not have time to discuss the implications
of sum rules in semileptonic decay and lattice QCD results.

In order to present the new developments in the theory of heavy quarks in their
proper context and to fully appreciate their significance I will briefly review
some of the key early work on heavy quark theory.

\noindent{\bf 2.  Heavy Quark Effective Theory}

The part of the QCD Lagrange density that contains a heavy quark $Q$ is
 $$
{\cal L} = \bar{Q} (i\Dslash - m_Q) Q~.\eqno (2)
 $$
For situations where the heavy quark $Q$ is interacting with light degrees of
freedom (i.e., light quarks and gluons) carrying momentum much less than its
mass, $m_Q$, it is appropriate to take the limit $m_Q \rightarrow \infty$ with
the heavy quark four-velocity, $v^\mu$, held fixed.$^1$  In this limit the
interactions of the heavy quark become independent of its mass and spin
resulting in the approximate heavy quark spin-flavor symmetries of QCD.

To take this limit write
 $$(x) = e^{-im_{Q}v\cdot x} h_v^{(Q)} (x)\eqno (3)
 $$
where
 $$
\vslash h_v^{(Q)} = h_v^{(Q)}\eqno (4)
 $$
Putting eq. (3) into (2) gives
 $$
{\cal L}_0 = \bar h_v^{(Q)} (i\Dslash + m_Q (\vslash - 1)) h_v^{(Q)}~.\eqno (5)
 $$
Using the constraint (4) this can be simplified to$^{2,3}$
 $$
{\cal L}_0 = \bar h_v^{(Q)} i v \cdot D h_v^{(Q)}~.\eqno (6)
 $$
Note that the Lagrange density in eq. (6) is independent of the heavy quark's
mass and it's spin.  Consequently the heavy quark effective theory has a spin
flavor symmetry.$^1$  For charm and bottom quarks moving with the same velocity
this is an SU(4) symmetry.  Much of the predictive power of the heavy quark
effective theory arises because of this symmetry.

The heavy quark field $h_v^{(Q)}$ destroys a quark $Q$ but it does not create
the corresponding antiquark.  Pair creation does not occur in the heavy quark
effective theory.

\noindent{\bf 3.  $1/m_Q$ Corrections}

The heavy quark effective theory in (6) represents the $m_Q \rightarrow \infty$
limit of QCD.  At finite $m_Q$ there are corrections suppressed by powers of
$1/m_Q$.  These can be included in a systematic fashion.  In general
 $$
Q(x) = e^{-im_{Q} v\cdot x} [h_v^{(Q)} (x) + \chi_v^{(Q)} (x)]\eqno (7)
 $$
where
 $$
\vslash h_v^{(Q)} = h_v^{(Q)} \quad {\rm and} \quad \vslash \chi_v = -
\chi_v^{(Q)}\eqno (8)
 $$
The equation of motion for the heavy quark field $Q$
 $$
(i \Dslash - m_Q) Q = 0 \eqno (9)
 $$
can be used to express $\chi_v^{(Q)} (x)$ in terms of $h_v^{(Q)} (x)$ order by
order in $1/m_Q$.  Putting (7) into (9) and using (8) gives
 $$
\chi_v^{(Q)} = {1\over 2m_Q}  i\Dslash [h_v^{(Q)} + \chi_v^{(Q)}] \eqno (10)
 $$
which implies that
 $$
\chi_v^{(Q)} = {1\over 2m_Q} i\Dslash h_v^{(Q)} + {\cal O} (1/m_Q^2)~. \eqno
(11)
 $$
Using this in eq. (7) and then plugging (7) into the Lagrange density (2) gives
the heavy quark effective theory including $1/m_Q$ corrections.
 $$
{\cal L} = {\cal L}_0 + {\cal L}_1 \eqno (12)
 $$
with ${\cal L}_0$ given by eq. (6) and$^{4,5}$
 $$
{\cal L}_1 = \bar h_v^{(Q)} {(iD)^2\over 2m_Q} h_v^{(Q)} - a_2 (\mu) \bar
h_v^{(Q)} g {G_{\alpha\beta}\sigma^{\alpha\beta}\over 4m_Q} h_v^{(Q)} \eqno
(13)
 $$
with $a_2(\mu) = 1$.  In eq. (13) $g$ is the strong gauge coupling and
$G_{\alpha\beta}$ is the gluon field strength tensor.  The procedure we have
outlined above amounts to matching tree graphs in QCD with those in the heavy
quark effective theory.  When loops are included $a_2$ develops subtraction
point dependence because the operator $\bar h_v^{(Q)} g G_{\alpha\beta}
\sigma^{\alpha\beta} h_v^{(Q)}$ requires renormalization.  In the leading
logarithmic approximation
 $$
a_2 (\mu) = [\alpha_s (m_Q)/\alpha_s (\mu)]^{9/(33-2 n_f)} \eqno (14)
 $$
where $n_f$ is the number of light quark flavors.

The first term in eq. (13) is the heavy quark kinetic energy.  It breaks the
heavy quark flavor symmetry but not the spin symmetry.  The second term in eq.
(13) is the energy from the interaction of the heavy quark's color magnetic
moment with the chromomagnetic field.  It breaks both the spin and flavor
symmetries.

\noindent{\bf 4.  Spectroscopy of Heavy Hadrons}

In the $m_Q \rightarrow \infty$ limit hadrons containing a single heavy quark
$Q$ are classified not only by their total spin $\vec S$ but also by the spin
of their light degrees of freedom$^6$
 $$
\vec S_\ell = \vec S - \vec S_Q ~. \eqno (15)
 $$
Since $s_Q = 1/2$, in this limit hadrons containing a single heavy quark occur
in degenerate doublets labelled by the spin of the light degrees of freedom
$s_\ell$ and with total spins
 $$
s = s_\ell \pm 1/2 \eqno (16)
 $$
An exception occurs for $s_\ell = 0$ where there is only one state with $s =
1/2$.  For mesons with $Q{\bar q}~(q = u ~or~ d)$ flavor quantum numbers the
ground state doublet has negative parity and $s_\ell = 1/2$ giving a doublet of
spin-zero and spin-one mesons.  For $Q = c$ they are the $D$ and $D^*$ mesons
and for $Q = b$ they are the $B$ and $B^*$ mesons.  In the $Q = c$ case an
excited doublet of positive parity mesons with $s_\ell = 3/2$ has been
observed.  The hadrons in this doublet are sometimes called $D^{**}$ mesons and
have total spins one and two.

Baryons with $Q{qq}$ flavor quantum numbers have also been observed.  The
ground state isospin zero baryons have positive parity and $s_\ell = 0$ and are
called $\Lambda_Q$ baryons.  The ground state $I = 1$ baryons have positive
parity and  $s_\ell = 1$ and come  in a doublet with $s = 1/2$ and $3/2$.  They
are called $\Sigma_Q$ and $\Sigma_Q^*$ baryons.  For $Q = c$ the $\Lambda_c$
and $\Sigma_c$  baryons have been observed and for $Q = b$ the $\Lambda_b$
baryon has been observed.  In the charm case two excited baryons have also been
observed.  Their properties are consistent with being a negative parity doublet
of $I = 0$ baryons with $s_\ell = 1$ giving total spins $1/2$ and $3/2$.

The mass of a hadron $H_Q$ containing a single heavy quark $Q$ can be expanded
in powers of $1/m_Q$.  Up to order $1/m_Q$ it has the form
$$\eqalign{
m_{{H}_{Q}} &= m_Q + \bar \Lambda - \langle H_Q | \bar{h}_v^{(Q)} {(iD)^2\over
2m_Q} h_v^{(Q)} | H_Q\rangle/2m_H \cr
&+ a_2 (\mu) \langle H_Q | \bar{h}_\sigma^{(Q)} {g
G_{\alpha\beta}\sigma^{\alpha\beta}\over 4m_Q} h_v^{(Q)} | H_Q\rangle/2m_H +
{\cal O} (1/m_Q^2)~.\cr} \eqno(17)
$$
The first term on the rhs of equation (17), $m_Q$, is the heavy quark pole
mass.  The second $\bar\Lambda$ is the mass of the light degrees of freedom in
the hadron.  It does not depend on the heavy quark mass but does   depend on
the quantum numbers of the light degrees of freedom.  The third term is the
heavy quark's kinetic energy and the final term is its chromomagnetic energy.
Only the last term depends on the spin of the heavy quark and it causes the
splittings in the hadron doublets mentioned earlier.  For example
$$
m_{B^{*}} - m_B = -{4\over 3} a_2 (\mu) \langle B| \bar h_v^{(b)}
{gG_{\alpha\beta}\over 4m_b} \sigma^{\alpha\beta} h_v^{(b)} | B \rangle/2m_B
{}~.\eqno(18)
$$

The heavy quark pole mass $m_Q$ is not a physical quantity and its perturbative
expansion has an infrared renormalon ambiguity of order
$\Lambda_{QCD}$.$^{7,8}$  Nonetheless, it is very convenient to introduce it.
As long as final expressions that are compared with experiment express physical
quantities in terms of other physical quantities the fact that the pole mass
itself is not really well defined is of no consequence.$^{9,10}$

\noindent{\bf 5. Exclusive $B \rightarrow D^{(*)} e\bar\nu_e$ Decay}

The rates for $B \rightarrow D e\bar\nu_e$ and $B \rightarrow D^* e \bar\nu_e$
are determined by the value of $|V_{bc}|$  and the hadronic matrix element of
the weak current $\bar c \gamma_\mu (1-\gamma_5)b$ between $B$ and $D^{(*)}$
states.  The application of heavy quark effective theory involves a two step
process.  First is matching the current $\bar c \gamma_\mu (1 - \gamma_5)b$
onto operators in the heavy quark effective theory.  In the leading logarithmic
approximation this matching takes the simple form$^{11}$
$$
\bar c \gamma_\mu (1 - \gamma_5) b = \left[{\alpha_s (m_b)\over \alpha_s
(m_c)}\right]^{-6/25} \left[{\alpha_s (m_c)\over \alpha_s (\mu)}\right]^{a_{L}}
\cdot \bar h_{v'}^{(c)} \gamma_\mu (1 - \gamma_5) h_v^{(b)}\eqno(19)$$
where
$$
a_L = {8\over 27} [v \cdot v' r (v \cdot v') -1]\eqno(20)$$
and
$$
r (v \cdot v') = {1\over\sqrt{(v \cdot v')^2 - 1}} \ln (v \cdot v' + \sqrt{(v
\cdot v')^2 -1}).\eqno(21)$$
Note that for $v \cdot v' \not= 1$ the coefficient of the current in the
effective theory $\bar h_{v'}^{(c)} \gamma_\mu (1-\gamma_5) h_v^{(b)}$ depends
on the subtraction point $\mu$.  In the effective theory where the charm and
bottom quarks are both treated as heavy the operator $\bar h_{v'}^{(c)}
\gamma_\mu (1-\gamma_5) h_v^{(b)}$ requires renormalization.  It's matrix
elements have a $\mu$ dependence that cancels that of its coefficient.
However, at zero recoil $v \cdot v' = 1$ the coefficient is independent of
$\mu$.  At this kinematic point the operator is the conserved current
associated with the spin-flavor symmetries of the heavy quark effective theory
and consequently it is not renormalized.

Matrix elements of $\bar h_{v'}^{(c)} \Gamma h_v^{(b)}$ in the heavy quark
effective theory between $B$ and $D^*$ states are related by heavy quark spin
symmetry to a single universal function of $v \cdot v'$,$^1$
$$
{\langle D(v') | \bar h_{v'}^{(c)} \Gamma h_v^{(b)} | B (v) \rangle\over
\sqrt{m_B m_D}} = \xi (v \cdot v') Tr \left\{{(\vslash' + 1)\over 2} \Gamma
{(\vslash + 1)\over 2} \right\}\eqno(22)$$
$$
{\langle D^* (v', \varepsilon) |\bar h_{v'}^{(c)} \Gamma h_v^{(b)} |
B(v)\rangle\over \sqrt{m_B m_{D^{*}}}} = \xi (v \cdot v') Tr
\left\{\varepsilon^* {(\vslash' +1)\over 2} \Gamma {(\vslash + 1)\over 2}
\gamma_5 \right\}~.\eqno(23)$$
For $v \cdot v' \not= 1$ the Isgur--Wise function $\xi (v \cdot v')$ depends on
the subtraction point $\mu$.  However, at zero recoil heavy quark flavor
symmetry fixes the normalization$^{1,12,13}$ of $\xi$,
$$
\xi (1) = 1.\eqno(24)$$
Equations (22) and (23) hold in the $m_{c,b} \rightarrow \infty$ limit.  In
general there are $\Lambda_{QCD}/m_{c,b}$ corrections.  However, at zero recoil
it has been shown that corrections first arise at order
$\Lambda_{QCD}^2/m_{c,b}^2$.$^{13,14}$  This important result opens an avenue
for the  precise determination of $|V_{cb}|$ from exclusive $B \rightarrow D^*
e \bar \nu_e$ decay.

Neglecting nonperturbative corrections, suppressed by powers of
$(\Lambda_{QCD}/m_{b,c})$, the zero recoil, the matrix elements of the axial,
and vector currents are
$$
{\langle D(v) | \bar c \gamma_\mu b | B(v) \rangle\over \sqrt{m_B m_D}} = 2
\eta_V v_\mu\eqno (25)$$
$$
{\langle D^* (v, \varepsilon) | \bar c \gamma_\mu b | B(v)\rangle\over
\sqrt{m_B m_{D^{*}}}} = 2 \eta_A \varepsilon_\mu^* ,\eqno (26)$$
where $\eta_V$ and $\eta_A$ are QCD correction factors from matching currents
in the full theory onto those in the effective theory.  In the leading
logarithmic approximation where $\ln (m_b/m_c)$ is treated as large and all
terms of order $[\alpha_s \ln (m_b/m_c)]^n$ are summed$^{15,16}$
$$
\eta_V = \eta_A = \left[{\alpha_s (m_b)\over \alpha_s (m_c)}\right]^{-6/25}
.\eqno (27)$$
However, since $m_b/m_c$ is not that large a better approximation is to keep
the  full dependence on $m_c/m_b$.  The coefficients $\eta_V$ and $\eta_A$ have
been calculated including two loop terms that come from vacuum polarization
insertions and are proportional to
$$
\beta^{(0)} = 11 - {2\over 3} n_f.\eqno (28)$$
The result is$^{13,17,18}$
$$
\eta_V = 1 + {1\over 3} {\bar{\alpha}_s (m_b)\over \pi}\phi (m_c/m_b) +
\left({\bar{\alpha}_s (m_b)\over \pi}\right)^2 \left[{1\over 72} \phi (m_c/m_b)
\beta^{(0)} +...\right] +...~~,\eqno (29)$$
and
$$\eta_A = 1 + {1\over 3} {\bar{\alpha}_s (m_b)\over \pi} [\phi (m_c/m_b)-2] +
\left({\bar{\alpha}_s (m_b)\over \pi}\right)^2 $$
$$
\cdot \left[\left({5\over 72} \phi (m_c/m_b)-{14\over 72}\right) \beta^{(0)}
+...\right] +...~~,\eqno (30)$$
where
$$
\phi(z) = - 3 \left({1 + z\over 1 -  z}\right) \ln z - 6 ~~,\eqno (31)$$
$m_c$ and $m_b$ are heavy quark pole masses and $\bar{\alpha}_s$ is the
$\overline{MS}$ strong coupling.  The ellipses in the square brackets are terms
independent of $n_f$.  There are reasons to believe that the order
$\bar{\alpha}_s^2 (m_b)$ piece proportional to $\beta^{(0)}$ provides a good
approximation to the full order $\bar{\alpha}_s^2 (m_b)$ term.  That is true
for $R(e^+e^- \rightarrow$ hadrons), $\Gamma(\tau \rightarrow \nu_\tau +$
hadrons) and the relation between the heavy quark pole mass $m_Q$ and the
running heavy quark  $\overline{MS}$ mass $\bar{m}_Q (m_Q)$:
$$ R(e^+ e^- \rightarrow {\rm hadrons}) = 3 \left(\sum_i Q_i^2\right) \Bigg[ 1
+ {\bar{\alpha}_s (\sqrt{s})\over \pi}$$
$$
+ (0.17 \beta^{(0)} + 0.08) \left({\bar{\alpha}_s (\sqrt{s})\over \pi}\right)^2
+ ... \Bigg]\eqno (32)$$
$$
{\Gamma(\tau \rightarrow \nu_\tau + {\rm hadrons})\over 3 \Gamma (\tau
\rightarrow \nu_\tau \bar{\nu}_e e)} = 1 + {\bar{\alpha}_s (m_\tau)\over \pi} +
(0.57 \beta^{(0)} + 0.08) \left({\bar{\alpha}_s (m_\tau)\over \pi}\right)^2
+...\eqno (33)$$
$$
m_Q/\bar m_Q (m_Q) = 1 + {4\over 3} {\bar{\alpha}_s (m_Q)\over \pi} + (1.56
\beta^{(0)} - 1.05) \left({\bar{\alpha}_s (m_Q)\over \pi}\right)^2 +
...~~.\eqno (34)$$
Evaluating eqs. (29) and (30) with $m_c/m_b = 0.30$ and $\bar{\alpha}_s (m_b) =
0.20$ gives
$$\eqalign{
\eta_V &= 1 + 0.02 + 0.004\cr
\eta_A &= 1 - 0.03 - 0.005~~,\cr} \eqno (35)$$
In eqs. (35) the second and third terms are the ones of order $\bar{\alpha}_s$
and $\bar{\alpha}_s^2 \beta^{(0)}$ respectively. Also we have taken $n_f = 2$
which gives $\beta^{(0)} = 9$.  Note that the two loop term is much smaller
than the one loop term indicating that the perturbation series is well behaved.

Nonperturbative corrections to (25) and (26) are of order
$(\Lambda_{QCD}/m_{c,b})^{n + 2} , n = 0,1,...~$.  For $n = 0$ these have been
characterized in terms of matrix elements of various operators in  the heavy
quark effective theory and estimated using phenomenological models.$^{19}$  In
addition the corrections to eqs. (25) and (26) that are enhanced by $\ln m_\pi$
or factors of $1/m_\pi$ have been computed using chiral perturbation theory.
These have an interesting form.$^{20}$  The correction of order
$(\Lambda_{QCD}/m_{c,b})^2$ is enhanced by $\ln m_\pi$ but corrections
suppressed by higher powers $(\Lambda_{QCD}/m_{c,b})^{n + 2}, n = 1,2,...$ are
enhanced by $(\Lambda_{QCD}/m_\pi)^n$.  Consequently, power suppressed terms
are important for all $n$. These corrections are calculable in terms of the
$D^* D\pi$ coupling.  Unfortunately the value of this coupling is not known.
This gives one of the major uncertainties in the size of the power correction
to eqs. (25) a!
nd (26).

\noindent{\bf 6. Inclusive $B \rightarrow X_{c,u} e \bar\nu_e$ Decay}

Over the past few years there has been great progress in our understanding of
inclusive semileptonic $B$ meson decay.$^{21,22,23,24}$  The strong interaction
physics relevant for this process is parametrized by the hadronic tensor
$$
W_{c,u}^{\mu\nu} = (2\pi)^3 \sum_X \delta^4 (p_B - q - p_X) \langle B |
J_{c,u}^{\mu\dagger} |X \rangle \langle X | J_{c,u}^\nu |B  \rangle\eqno (36)$$
and
$$
J_c^\mu = \bar c \gamma^\mu (1 - \gamma_5) b\eqno (37)$$
$$
J_u^\mu = \bar u \gamma^\mu (1 - \gamma_5) b\eqno (38)$$
$W^{\mu\nu}$ can be expanded in terms of scalar form factors $W_n, n = 1,2,...,
5$ that are functions of $q^2$ and $v\cdot q$.
$$W^{\mu\nu} = - g^{\mu\nu} W_1 + v^\mu v^\nu W_2 - i
\varepsilon^{\mu\nu\alpha\beta} v_\alpha q_\beta W_3$$
$$
+q^\mu q^\nu W_4 + (q^\mu v^\nu + q^\nu v^\mu) W_5~.\eqno (39)$$
The form factors $W_j$ are the imaginary parts of form factors that occur in
the matrix element of the time ordered product of weak currents.
$$
T_{c,u}^{\mu\nu} = - i \int d^4 x e^{-iq \cdot x} \langle B| T(
J_{c,u}^{\mu\dagger} (x) J_{c,u}^\nu (0) | B \rangle\eqno (40)$$
can be expanded in terms of scalar form factors
$$T^{\mu\nu} = - g^{\mu\nu} T_1 + v^\mu v^\nu T_2 -
i\varepsilon^{\mu\nu\alpha\beta} v_\alpha  q_\beta T_3$$
$$
+ q^\mu q^\nu T_4 + (q^\mu v^\nu + q^\nu v^\mu) T_5\eqno (41)$$
and
$$
Im T_{c,u} = - \pi W_{c,u}.\eqno (42)$$
Predictions for the form factors $T_j$ can be made by performing an operator
product expansion and making a transition to the heavy quark effective theory.
The leading operator encountered is $\bar b \gamma_\mu b$ and its matrix
element is known since it is the conserved b-quark number current.  Here there
is no need to make the transition to the heavy quark effective theory to
understand the $m_b$ dependence.  There are no dimension four operators and the
dimension 5 operators that occur are the $b$-quark kinetic energy and the
chromomagnetic dipole term that occur in ${\cal L}_1$ of eq. (13).
Consequently at leading order in $1/m_b$ the differential decay rate $d\Gamma/
dq^2 d E_e$ for inclusive semileptonic $B$-decay is given by free $b$-quark
decay.  There are no nonperturbative corrections of order
$(\Lambda_{QCD}/m_b)$.  Nonperturbative corrections of order
$(\Lambda_{QCD}/m_b)^2$ are characterized by the two dimensionless parameters
$$
K_b = - \langle B | \bar h_v^{(b)} {(iD)^2\over 2m_b^2} h_v^{(b)} | B
\rangle/2m_B\eqno (43)$$
$$
G_b = a (\mu) \langle B | \bar h_v^{(b)} {g G_{\mu\nu}\over 4m_b^2}
\sigma^{\mu\nu} h_v^{(b)} | B \rangle/2m_B~.\eqno (44)$$
Including perturbative corrections and nonperturbative corrections suppressed
by $(\Lambda_{QCD}/m_b)^2$ the $B \rightarrow X_c e \bar\nu_e$ semileptonic
decay rate is
$$\Gamma(B \rightarrow X_c e\bar\nu_e) = \Gamma_0 [(1 - 8\rho + 8 \rho^3
-\rho^4 -12\rho^2 \ln \rho)\eta_{incl}$$
$$+K_b (-1 + 8\rho - 8\rho^3 +\rho^4 + 12\rho^2 \ln \rho) + G_b (3 - 8\rho +
24\rho^2 -24\rho^3$$
$$
+5\rho^4 + 12\rho^2 \ln \rho)]\eqno (45)$$
where
$$
 \rho = m_c^2/m_b^2, \eqno (46)$$
and
$$
\Gamma_0 = {|V_{cb}|^2 G_F^2 m_b^5\over 192\pi^3}~.\eqno (47)$$
In $\Gamma_0 ~m_b$ is the $b$-quark pole mass and $\eta_{incl}$ gives the
effects of perturbative QCD corrections.   Results for $B \rightarrow X_u
e\bar\nu_e$ are obtained by taking $\rho = 0$ and $V_{cb} \rightarrow V_{ub}$.
$\eta_{incl}$ depends on $m_c/m_b$ so the perturbative QCD corrections are
different for $B \rightarrow X_c e\bar\nu_e$ and $B \rightarrow X_u e\bar\nu_e$
decay.  The nonperturbative corrections are quite small.  Furthermore, $G_b$ is
known from the $B^* - B$ mass splitting so the only uncertainty in the
nonperturbative corrections comes from the size of $K_b$. In eqs. (45)-(47)
$m_c$ and $m_b$ are the charm and bottom quark pole masses.   If $m_c$ is
eliminated by
$$
m_B - m_D = m_b - m_c + m_b K_b - m_c K_c + m_b G_b - m_c G_c ,\eqno (48)$$
then the decay rate is not too sensitive to the value of $m_b$.  For example,
as $m_b$ varies between 5GeV and 4.5GeV the rate $\Gamma (B \rightarrow  X_c e
\bar{\nu}_e)$ changes by only 20\%.

Neglecting the nonperturbative corrections the $B$ decay rate equals the
$b$-quark decay rate.  The perturbative QCD corrections of order
$\bar{\alpha}_s (m_b)$ have been computed and those of order $\bar{\alpha}_s
(m_b)^2$ proportional to $\beta^{(0)}$ are also known.  We write
$$\eqalign{
\eta_{incl} &= 1 - \left({\bar{\alpha}_s (m_b)\over \pi} \right) {2\over 3}
\left(\pi^2 - {25\over 4} + \delta_1 (m_c/m_b)\right)\cr
&- \left({\bar{\alpha}_s (m_b)\over \pi}\right)^2 (\beta^{(0)} \chi_\beta
(m_c/m_b) + ...) + ...~~.\cr} \eqno (49)$$
The function $\delta_1 (x)$ is known analytically.$^{25}$  It takes into
account the effect of the charm quark mass on the order $\bar{\alpha}_s$ QCD
corrections;  $\delta_1 (0) = 0$.  Numerically $\delta_1 (0.3) = - 1.11$.  The
function $\chi_ \beta (x)$ has been determined numerically yielding $\chi_\beta
(0) = 3.2$ and $\chi_\beta (0.3) = 1.7$.  Using $\bar{\alpha}_s (m_b) = 0.20$
and $m_c/m_b = 0.30$ gives$^{26}$
$$
\eta_{incl} = 1 - 0.11 - 0.06 + ...\eqno (50)$$
for $B \rightarrow X_c e \bar{\nu}_e$ decay and (using $m_c/m_b = 0$)
$$
\eta_{incl} = 1 - 0.15 - 0.11 + ...~~,\eqno (51)$$
for $B \rightarrow X_u e\bar{\nu}_e$ decay.  The second and third terms in
eqs. (50) and (51) are the pieces of order $\bar{\alpha}_s (m_b)$ and
$\bar{\alpha}_s (m_b)^2 \beta^{(0)}$ respectively.  In the ``two loop'' term we
have taken $n_f = 2$ which gives $\beta^{(0)} = 9$.  For $B \rightarrow X_u e
\bar{\nu}_e$ the perturbative series is not well behaved and the situation for
$B \rightarrow X_c e\bar{\nu}_e$ is somewhat marginal.  For inclusive
semileptonic $D \rightarrow X_s \bar{e} \nu_e$ decay similar formulas hold.
The perturbative QCD corrections can be deduced from eq. (49) with
$\bar{\alpha}_s (m_b) \rightarrow \bar{\alpha}_s (m_c)$ and $m_c/m_b
\rightarrow m_s/m_c \simeq 0$.  Here the QCD corrections are also not under
control.

The methods outlined above for inclusive semileptonic $B$ decay can also be
applied to nonleptonic $B$-decay.  Here one runs into a potential conflict
between the measured semileptonic branching ratio and the measured charm
multiplicity.$^{27,28}$  For the decays that come from $b \rightarrow c \bar{c}
s$ the charm quark masses take up most of the available energy.  Therefore, it
is not clear that local duality can be used to relate the quark level decay to
the hadron decay.  Furthermore, the perturbative QCD corrections in the quark
level decay may not be under control.  To accommodate the measured semileptonic
branching ratio$^{29}$
$B_{SL} = (10.4 \pm 0.4)\%$ requires about 40\% of the nonleptonic $B$ decays
to come from the $b \rightarrow c \bar{c} s$ mechanism.  This implies a charm
multiplicity $\langle n_c \rangle \simeq 1.3$.  However, the measured charm
multiplicity$^{30}$ is only $\langle n_c \rangle_{\exp} = 1.04 \pm 0.07$.  It
will take more data to resolve this issue.

\noindent{\bf 7.  The End Point Region of the Electron Spectrum}

The maximum electron energy in the exclusive decay $B \rightarrow X e
\bar{\nu}_e$ is
$$
E_e^{\max} = {m_B^2 - m_X^2\over 2m_B} ~.\eqno (52)$$
Therefore, semileptonic $B$ decays with electron energies greater than $(m_B^2
- m_D^2)/2m_B$ must have come from a $b \rightarrow u$ transition.  This
endpoint region of the electron energy spectrum is very important.
Understanding it in a model independent way may lead to a precise determination
of $V_{ub}$.

For inclusive $B \rightarrow X_u e \bar{\nu}_e$ decay the electron energy
spectrum, including nonperturbative effects of order $(\Lambda_{QCD}/m_b)^2$,
has been found using the operator product expansion methods outlined in the
previous section.  Neglecting perturbative QCD corrections$^{22,23}$
$$\eqalign{
{1\over\Gamma_0} {d\Gamma\over dy} &= \Bigg[2 (3-2y)y^2 -{20\over 3} y^3 K_b
- \left( 8 + {20\over 3} y \right)y^2 G_b\Bigg] \theta (1 - y) \cr
&+ {2\over 3} [K_b + 11G_b] \delta (1 - y) + {2\over 3} K_b \delta' (1 -
y)~,\cr} \eqno (53)$$
where
$$
y = (2E_e/m_b)~,~~~~~~~~~~~~~~~~~~~~~~~~~~~~~~~~~~~~~~~~~~~~\eqno (54)$$
and $K_b$ and $G_b$ are given in eqs. (43) and (44).  These matrix elements are
of order $\varepsilon^2$ where,
$$
\varepsilon = \Lambda_{QCD}/m_b~.\eqno (55)$$
The maximum electron energy for $b$-quark decay is $y = 1$ (i.e., $E_e =
m_b/2$).  However,  nonperturbative effects (e.g., motion of the $b$-quark  in
the $B$-meson) extend the maximum electron energy for $B$-meson decay  beyond
this point.  Since we are treating such effects as a power series in
$\varepsilon$ they are represented by singular terms at $y = 1$.  To all orders
in $\varepsilon$ the decay spectrum obtained from the operator product
expansion has the structure$^{31}$ (at zero'th order in $\alpha_s (m_b))$
$${1\over\Gamma_0} {d\Gamma\over dy} = \theta (1 - y) (\varepsilon^0 +
0\varepsilon + \varepsilon^2 + ...)$$
$$+\delta (1 - y) (0\varepsilon + \varepsilon^2 + ...) + \delta^{(1)} (1 - y)
(\varepsilon^2 + \varepsilon^3 + ...) + ...$$
$$
+ \delta^{(n)} (1 - y)  (\varepsilon^{n+1} + \varepsilon^{n+2} + ...) + ...
{}~~,\eqno (56)$$
where $\varepsilon^n$ denotes a term of that order, which may include a smooth
function of $y$.  In eq. (56) $\delta^{(n)}(1-y)$ denotes the n'th derivative
of $\delta(1-y)$ with respect to $y$.  The contribution to the total decay rate
of a term in $d\Gamma/dy$ of order $\varepsilon^n \delta^{(n)}(1-y)$ is of
order $\varepsilon^n$.

The semileptonic decay width for $b \rightarrow u$ is difficult to measure
because of background contamination from the dominant $b \rightarrow c$
semileptonic decays.  It is therefore, important to be able to compute the rate
in the endpoint region near $y = 1$.  One way to calculate the endpoint
spectrum is to weight the differential decay distribution $d\Gamma/dy$ in eq.
(56) by a normalized function of width $\sigma$ around $y = 1$.  We refer to
this process as smearing.  Most of the details of the smearing procedure are
unimportant; the only quantity of relevance is the width $\sigma$ of  the
smearing region.

The singular distribution $\varepsilon^m \delta^{(n)} (1-y)$ (where $m > n$)
smeared over a region of width $\sigma$ gives a contribution of order
$\varepsilon^n/\sigma^{n+1}$ to $d\Gamma/dy$.  If the width $\sigma$ of the
smearing region is of order $\varepsilon^p$ the generic term $\varepsilon^m
\delta^{(n)} (1 - y)$ yields a contribution of order $\varepsilon^{m -
(n+1)p}$.  Since $m > n$ this shows that  the $1/m_b$ expansion for the
spectrum breaks down unless $p \leq 1$, i.e., the smearing region cannot be
narrower than $\varepsilon$.    The divergence for $p>1$ is not associated with
the failure of the operator product expansion due to resonances with masses of
order the QCD scale.  The region of the electron energy spectrum for which such
resonances dominate the final hadronic states is of width $\varepsilon^2$,
while the expansion breaks down upon smearing over any region of size
$\varepsilon^{1 + \delta}$, where $\delta > 0$.

If the smearing region is chosen of order $\varepsilon$ the form of the
expansion in eq. (56) shows that the leading terms of the form $\theta (1-y)$
and  $\varepsilon^{n+1} \delta^{(n)} (1-y)$ all contribute at order unity to
the smeared spectrum.  Thus one can obtain the decay spectrum smeared over a
width $\varepsilon$ if the leading singularities can be summed.  The sum of the
leading singularities produces a distribution $d\Gamma/dy$ of width
$\varepsilon$ and height of order unity (i.e., of the same order as the free
quark distribution).  Neubert and Bigi, et al.,  have shown how to sum the
leading singularities.$^{32,33}$  They are characterized by the matrix elements
$$
{1\over 2m_B} \langle B| \bar{h}_v^{(b)} iD^{\mu_{1}}...~iD^{\mu_{n}} h_v^{(b)}
|B\rangle = A_n v_{\mu_{1}}... v_{\mu_{n}} + ...~~.\eqno (57)$$
The ellipsis on the right side of eq. (57) denote other Lorentz structures.
For example, with $n = 2$ the matrix element is,
$$
{1\over 2m_B} \langle B | \bar{h}_v^{(b)} iD^{\mu_{1}} iD^{\mu_{2}} h_v^{(b)}
|B \rangle = A_2 (v_{\mu_{1}} v_{\mu_{2}} - g_{\mu_{1} \mu_{2}}) ~~,\eqno
(58)$$
since $v \cdot D h_v^{(b)} = 0$.  Contracting on $\mu_1$ and $\mu_2$ gives
$$
A_2 = {2\over 3} m_b K_b ~.\eqno (59)$$
Heavy quark symmetry implies that $A_0 = 1$ and the equation of motion $v \cdot
Dh_v^{(b)} = 0$ implies that $A_1 = 0$.  The quantities $A_n$ have dimensions
of mass to the power $n$. In terms of them the sum of the leading singularities
in the electron spectrum is characterized by a shape function $S(y)$
$$
{1\over\Gamma_0} {d\Gamma\over dy} = 2y^2 (3 - 2y) S(y)\eqno (60)$$
$$
S(y) = \sum_{n = 0}^\infty {(-1)^n\over n!} \left({A_n\over m_b^n}\right)
{d^n\over dy^n} \theta (1-y)~~.\eqno (61)$$
Perturbative QCD corrections are also singular in the endpoint region.  Summing
the leading perturbative QCD singularities (i.e., the Sudakov double
logarithms)  changes the shape function to$^{31}$
$$
S(y) = \sum_{n=0}^\infty {(-1)^n\over n!} \left({A_n\over m_b^n}\right)
{d^n\over dy^n} R(y)\eqno (62)$$
where
$$
R(y) = \exp \left[ -{2\over 3\pi} \alpha_s \ln^2 (1-y) \right]~~.\eqno (63)$$
Recently Korchemsky and Sterman have shown how to sum all the singular
perturbative QCD corrections.$^{34}$

Unfortunately the quantities $A_n$ are not known.  However, the same quantities
characterize the endpoint photon spectrum in $B \rightarrow X_s \gamma$.  So
there is hope that a detailed study of the photon spectrum in $B \rightarrow
X_s \gamma$ will determine the endpoint region of the electron spectrum in $B$
decays.$^{33,34,35}$

The methods outlined in this section for describing the endpoint region of the
electron spectrum apply when this region is dominated by many states with
masses of order $\sqrt{m_b \Lambda_{QCD}}$.  In the ISGW$^{36}$ model the
endpoint region where $b \rightarrow c$ transitions are forbidden is dominated
by the single decay mode $B \rightarrow \rho e \bar{\nu}_e$.  If $\rho$
dominance is found to hold experimentally then the sum of the leading
singularities is not a valid description of a region of electron energy which
is as small as the difference between the $B \rightarrow X_u e \bar{\nu}_e$ and
$B \rightarrow X_c e \bar{\nu}_e$ end points.

If the endpoint region is dominated by the rho meson there are other avenues
available to determine $V_{ub}$.  For example, exclusive $B$ and $D$ decays can
be used.  For $D \rightarrow \rho \bar{e} \nu_e$ the weak mixing angles are
known and the form factors for this decay mode to determine them for $B
\rightarrow \rho e \bar{\nu}_e$.  Using heavy quark symmetry and isospin
symmetry$^{37}$
$$
\langle \rho (k) | \bar{u} \gamma_\mu (1 - \gamma_5) b|B\rangle / \sqrt{2m_B}
= \left({\alpha_s (m_b)\over \alpha_s (m_c)}\right)^{-6/25} \langle \rho
(k)|\bar{d} \gamma_\mu (1 - \gamma_5) c|D\rangle/\sqrt{2m_D} ~.\eqno (64)$$
In the above perturbative QCD effects have been included in the leading
logarithmic approximation.  If light quark SU(3) symmetry is applied instead of
isospin symmetry then the decay $D \rightarrow K^*  \bar{e} \nu_e$ can be used
(instead of the Cabibbo suppressed decay $D \rightarrow \rho \bar{e} \nu_e$).
The form factors for this decay have already been measured.  Some problems with
this approach are the presence of $1/m_{c,b}$ corrections and possibly large
higher order perturbative QCD corrections.$^{38}$

\noindent{\bf 8.  References}

\item{1}  N. Isgur and M. Wise, {\it Phys. Lett.} {\bf B232} (1989) 113; {\bf
B237} (1990) 527.
\item{2}  E. Eichten and B. Hill, {\it Phys. Lett.} {\bf B234} (1990) 511.
\item{3}  H. Georgi, {\it Phys. Lett.} {\bf B240} (1990) 447.
\item{4}  E. Eichten and B. Hill, {\it Phys. Lett.} {\bf B243} (1990) 427.
\item{5}  A. Falk, et al., {\it Nucl. Phys.} {\bf B357} (1991) 185.
\item{6}  N. Isgur and M. Wise, {\it Phys. Rev. Lett.} {\bf 66} (1991) 1130.
\item{7}  I. Bigi, et al., TPI-MINN-94/4-T (1994).
\item{8}  M. Beneke, et al., MPI-PhT/94-18 (1994).
\item{9}  M. Luke, et al., UTPT-94-21 (1994).
\item{10} M. Neubert and C. Sachrajda, CERN-TH 7312/94 (1994).
\item{11} A. Falk, et al., {\it Nucl. Phys.} {\bf B343} (1990) 1.
\item{12} S. Nussinov and W. Wetzel, {\it Phys. Rev.} {\bf D36} (1987) 130.
\item{13} M. Voloshin and M. Shifman, {\it Sov. J. Nucl. Phys.} {\bf 47} (1988)
511.
\item{14} M. Luke, {\it Phys. Lett.} {\bf B252} (1990) 447.
\item{15} M. Voloshin and M. Shifman, {\it Sov. J. Nucl. Phys.} {\bf 45} (1987)
292.
\item{16} H. Politzer and M. Wise, {\it Phys. Lett.} {\bf B206} (1988) 681;
{\bf B208} (1988) 504.
\item{17} J.E. Paschalis and G. Gounaris, {\it Nucl. Phys.} {\bf B222} (1983)
473; F. Close, et al., {\it Phys. Lett.} {\bf B149} (1984) 209.
\item{18} M. Neubert, CERN-TH 7454/94 (1994).
\item{19} For a review of QCD sum rule calculations of $1/m_Q$ corrections
see: M. Neubert, {\it Phys. Rep.} {\bf 245} (1994) 261.
\item{20} L. Randall and M. Wise, {\it Phys. Lett.} {\bf B303} (1993) 135.
\item{21} J. Chay, et al., {\it Phys. Lett.} {\bf B247} (1990) 399.
\item{22} M. Voloshin and M. Shifman, {\it Sov. J. Nucl. Phys.} {\bf 41} (1985)
120; I. Bigi, et al., {\it Phys. Lett.} {\bf B293} (1992) 430; B. Blok, et al.,
{\it Phys. Rev.} {\bf D49} (1994) 3356; I. Bigi, et al., {\it Phys. Rev. Lett.}
{\bf 71} (1993) 496.
\item{23} A. Manohar and M. Wise, {\it Phys. Rev.} {\bf D49} (1994) 1310.
\item{24} T. Mannel, {\it Nucl. Phys.} {\bf B413} (1994) 396.
\item{25} M. Jezabek and J.S. Kuhn, {\it Nucl. Phys.} {\bf B314} (1989) 1.
\item{26} M. Luke, et al., UTPT 94-24 (1994); UTPT 94-27 (1994).
\item{27} I. Bigi, et al., {\it Phys. Lett.} {\bf B323} (1994) 408; A. Falk, et
al., CALT-68-1933 (1994).
\item{28} E. Bagan, et al., DESY 94-172 (1994).
\item{29} M. Aguilar--Benitez, et al., (Particle Data Group) {\it Phys. Rev.}
{\bf D50} (1994) 1173.
\item{30} P. Roudeau, {\it Heavy Quark Physics} (Rapporteur Talk at ICHEP 94,
Glasgow, July 20-27, 1994)
\item{31} A. Falk, et al., {\it Phys. Rev.} {\bf D49} (1994) 4553.
\item{32} M. Neubert, {\it Phys. Rev.} {\bf D49} (1994) 3392.
\item{33} I. Bigi, et al., CERN-TH 7129/93 (1993).
\item{34} G. Korchemsky and G. Sterman, ITP-SB-94-35 (1994).
\item{35} M. Neubert, {\it Phys. Rev.} {\bf D49} (1994) 4623.
\item{36} N. Isgur, et al., {\it Phys. Rev.} {\bf D39} (1989) 799; D. Scora and
N. Isgur, CEBAF-TH-94-14 (1994).
\item{37} N. Isgur and M. Wise, {\it Phys. Rev.} {\bf D42} (1990) 2388.
\item{38} D. Broadhurst and A. Grozin, OUT-4102-52 (1994).

\bye